\documentclass[%
 reprint,%
 amssymb, amsmath,%
 aip,cha%
]{revtex4-1}
\usepackage[columnwise]{lineno}

\usepackage{docs}%
\usepackage{bm}%
\usepackage[colorlinks=true,linkcolor=blue]{hyperref}%
\expandafter\ifx\csname package@font\endcsname\relax\else
 \expandafter\expandafter
 \expandafter\usepackage
 \expandafter\expandafter
 \expandafter{\csname package@font\endcsname}%
\fi
\hypersetup{
	colorlinks=true,
	linkcolor=blue,
	filecolor=magenta,}
\usepackage{graphicx}
\usepackage{lipsum}

\begin{document}

\title{Effect of spin glass frustration on exchange bias in NiMn/CoFeB bilayers.} %

\author{Sagarika Nayak}%
\affiliation{Laboratory for Nanomagnetism and Magnetic Materials (LNMM), School of Physical Sciences, National Institute of Science Education and Research (NISER), HBNI, P.O.- Jatni, 752050, India}%
\author{Palash Kumar Manna}%
\affiliation{Laboratory for Nanomagnetism and Magnetic Materials (LNMM), School of Physical Sciences, National Institute of Science Education and Research (NISER), HBNI, P.O.- Jatni, 752050, India}%
\author{Braj Bhusan Singh}%
\affiliation{Laboratory for Nanomagnetism and Magnetic Materials (LNMM), School of Physical Sciences, National Institute of Science Education and Research (NISER), HBNI, P.O.- Jatni, 752050, India}%
\author{Subhankar Bedanta}%
\email{sbedanta@niser.ac.in}
\affiliation{Laboratory for Nanomagnetism and Magnetic Materials (LNMM), School of Physical Sciences, National Institute of Science Education and Research (NISER), HBNI, P.O.- Jatni, 752050, India}%
\date{October 2020}%


\begin{abstract}
Exchange bias in ferromagnetic/antiferromagnetic systems can be explained in terms of various interfacial phenomena. Among these spin glass frustration can affect the magnetic properties in exchange bias systems. Here we have studied a NiMn/CoFeB exchange bias system in which spin glass frustration seems to play a crucial role. In order to account the effect of spin glass frustration on magnetic properties, we have performed the temperature and cooling field dependence of exchange bias. We have observed the decrease of exchange bias field ($H_{EB}$) with cooling field ($H_{FC}$) whereas there is not significant effect on coercive field ($H_{C}$). Exponential decay of $H_{EB}$ and $H_{C}$ is found in these exchange bias systems. Further, training effect measurements have been performed to study the spin relaxation mechanism. We have fitted the training effect data with frozen and rotatable spin relaxation model. We have investigated the ratio of relaxation rate of interfacial rotatable and frozen spins in this study. The training effect data are also fitted with various other models. Further, we observed the shifting of peak temperature towards higher temperature with frequency from the ac susceptibility data. 
\end{abstract}

\maketitle
\section{Introduction}
Historically, exchange bias was first studied in ferromagnetic (FM)/antiferromagnetic (AFM) systems~\cite{Xi-2002, Chi-2020, Zhang-2001, Wu-1998, Zhong-2014}. However, study of exchange bias in FM/spin-glass (SG) systems present a unique opportunity to explore how the SG component can also induce a unidirectional anisotropy leading to exchange bias. Exchange bias effects have been investigated in spin-glass AgMn, CuMn dilute alloys and also in FM/SG nanocomposite systems~\cite{Ali-2007,Yuan-2010}.  

Mn based AFM materials like FeMn, IrMn, PtMn, PdMn, NiMn have a good thermal stability that is the main requirement of exchange biased spintronic devices~\cite{Dai-2003}. One other important property of these above AFM’s is their high  N\'{e}el temperature $T_{N}$. For example, the $T_{N}$ of $Ni_{50}$$Mn_{50}$ is 1070 K~\cite{Baltz-2018}. As-grown NiMn is paramagnetic with a face centred cubic (FCC) structure whereas post deposition annealing develops antiferromagnetic phase with face centred tetragonal (FCT) structure~\cite{Groudeva-Zotova-2003,Akbulut-2016}. The lattice constants of FCT NiMn are a = b = 3.74 Å and c = 3.52 Å ~\cite{Groudeva-Zotova-2003,Akbulut-2016}. EB properties was reported in Co/NiMn system deposited on Cu (001) substrate. EB has also been investigated in some of the AFM’s such as NiMn, IrMn, $Cr_{2}$$O_{3}$ grown on Pt buffer layer~\cite{Akbulut-2016}. One can tune the microstructure by selecting the proper seed layer and hence the magnetic properties~\cite{Groudeva-Zotova-2003}. Exchange anisotropy at the FM/AFM interface can be controlled through proper selection of substrate, buffer layer and growth conditions due to the modification of structural orientations~\cite{Akbulut-2016,Lin-1994,Yang-1999}. AFM order can also be induced by choosing proper seed layer.  

In this study, we have performed the temperature and cooling field dependence of exchange bias to investigate the effect of interface and the SG $\lq${bulk}$\rq$ on magnetic properties. Contribution of $\lq${bulk}$\rq$ spins of NiMn has also been investigated from the variation of exchange bias field with its thickness. Besides, the increase in the relaxation rate ratio of rotatable and frozen spins indicates that not only the interface but also $\lq${bulk}$\rq$ spins of NiMn contribute to the observed results.
\begin{figure*}[hbt!]
	\centering
	\includegraphics[scale=0.6]{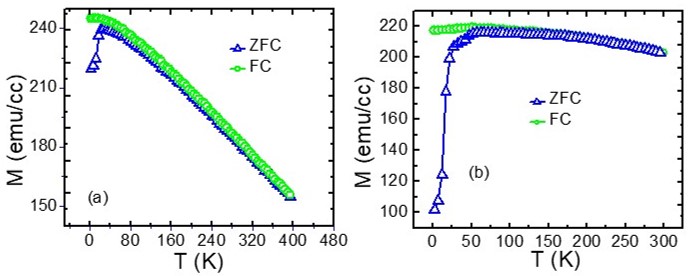}
	\caption{(a)-(b) $\it{M}$ vs $\it{T}$ curves under ZFC, FC condition and –d($M_{FC}$-$M_{ZFC}$)/dT vs $\it{T}$ plots for the samples S1 and S2.}
	\label{fig:fig-1}
\end{figure*}

\section{Experimental details:}
Deposition of $Ni_{50}$$Mn_{50}$ ($\it{t}$ = 5, 15 nm)/$Co_{40}$$Fe_{40}$$B_{20}$ (5 nm) bilayers on Si (100) substrate has been performed by dc magnetron sputtering at room temperature (RT). Pt has been deposited on top of Ta as seed layer. Ta has also been deposited as capping layer to avoid oxidation. We have not performed post deposition annealing to induce AFM order in NiMn. However, Pt and Ta are deposited to promote AFM order ~\cite{Akbulut-2016} and modify interfacial exchange coupling through microstructure adjustment. The substrate was rotated at 20 rotation per minute (rpm) speed during deposition of all the magnetic layers to avoid growth induced uniaxial magnetic anisotropy. The rate of deposition of Ta, Pt, NiMn, and CoFeB are 0.13, 0.3, 0.14, and 0.16 Å/sec, respectively. The magnetic measurements have been performed using a Quantum Design SQUID magnetometer (MPMS 3). All the sample details are given in table 1.

\begin{table*}
\small
  \caption{\ Details of sample nomenclature and configuration.}
  \label{tbl:example}
  \begin{tabular*}{\textwidth}{@{\extracolsep{\fill}}ll}
    \hline
    Sample name&Sample structure \\
    \hline
    S1 & Si(100)/Ta(3 nm)/Pt(2.5 nm)/NiMn (5 nm)/CoFeB(5 nm)/Ta (3 nm)\\
	S2 & Si(100)/Ta(3 nm)/Pt(2.5 nm)/NiMn (15 nm)/CoFeB(5 nm)/Ta (3 nm)\\
    \hline
  \end{tabular*}
\end{table*}

\section{Results and discussion:}
\subsection{Magnetization vs temperature}
To elucidate the magnetic nature of the films dc-magnetization measurements were performed as a function of temperature and fields. Figure 1 (a),(b) represent the magnetization ($\it{M}$) vs temperature ($\it{T}$) curves under zero field cooled (ZFC) and field cooled (FC) conditions in a magnetic field of 10 mT.  We found the peak temperatures ($T_{P}$) of 22, and 47 K in samples S1 and S2, thus, $T_{P}$ increases with the thickness of NiMn. The $T_{P}$ has broadened for higher thicknesses of NiMn due to increase in particle size and broader size distribution \cite{AnilKumar-2012}. We found the bifurcation in ZFC and FC $\it{M-T}$ curves in all the samples at a temperature known as irreversibility temperature ($T_{irr}$). 
\begin{figure*}[hbt!]
	\centering
	\includegraphics[scale=0.6]{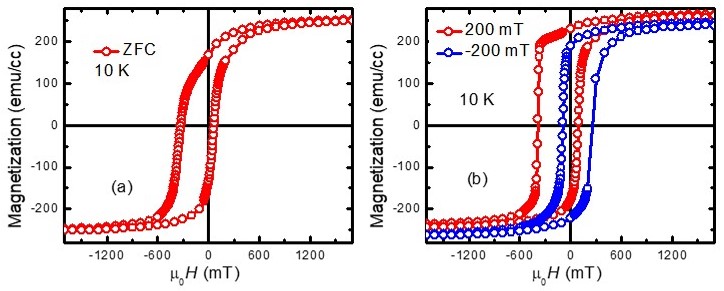}
	\caption{Plots of hysteresis loop of sample S2 for negative, positive field cooling (FC) in 2 kOe field (b) and zero field cooling (ZFC) (a) states, respectively.}
	\label{fig:figure2}
\end{figure*}
\subsection{Magnetic hysteresis}
Figure 2 shows the magnetic hysteresis loops of sample S2 in zero field cooled (ZFC), positive and negative field cooled conditions (200 mT field), respectively at 10 K temperature. We found the negative exchange bias of 12.78 mT strength in ZFC condition indicating that the presence of positive exchange interaction at the interface. We found the negative and positive exchange bias in this sample under positive and negative field cooled conditions. This indicates the presence of positive and negative exchange interaction at the interface. However, the magnitude of exchange bias is different for field cooling in 200 mT (14.66 mT) and -200 mT (8.29 mT) fields, respectively. This, indicates that the coupling strength is different under positive and negative field cooling conditions, respectively.  
\subsection{Temperature dependence of exchange bias}
\begin{figure*}[hbt!]
	\centering
	\includegraphics[scale=0.6]{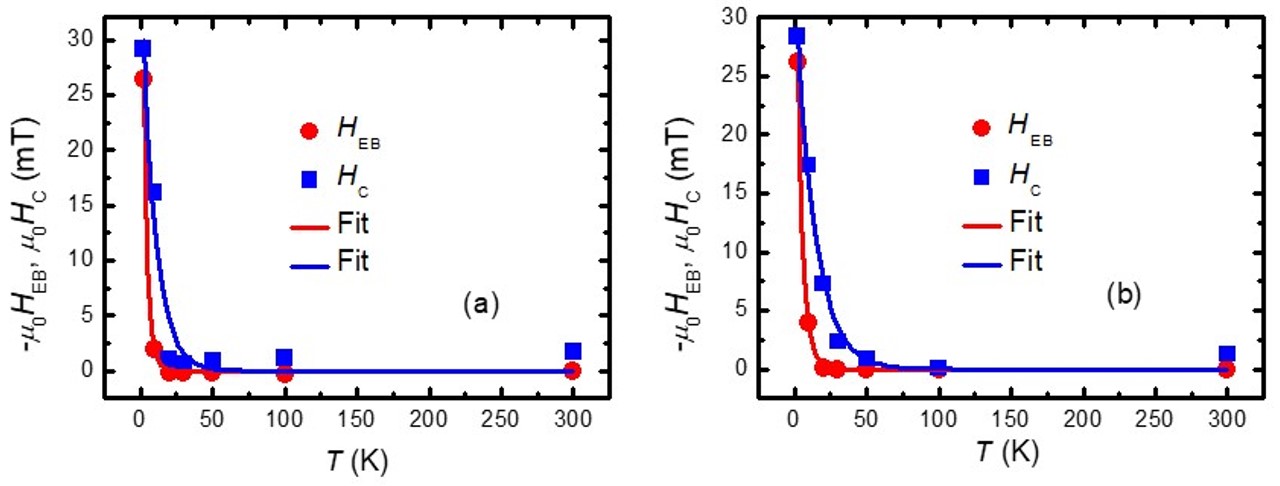}
	\caption{(a)-(b) Plots of $H_{EB}$, $H_{C}$ vs $\it{T}$ experimental data along with their fits using eqs. (1) and (2) for samples S1 and S2, respectively.}
	\label{fig:figure2}
\end{figure*}

\begin{table*}
\small
	\caption{\label{tab:table1}Fitting parameters obtained from the fits using eqs. (1) and (2), respectively.}
	\begin{tabular*}{\textwidth}{@{\extracolsep{\fill}}ll}
		\begin{tabular}{cccccc}
		    \hline 
			Sample name & $H_{EB}^0$ (mT) & $T_{1}$ (K) & $H_{C}^0$ (mT) & $T_{2}$ (K)\\
			\hline
			S1 & 51.18 $\pm$ 1.56 & 3.02 $\pm$ 0.13 & 36.91 $\pm$ 3.68 & 9.57 $\pm$ 1.70 \\
			S2 & 42.22 $\pm$ 0.50 & 4.16 $\pm$ 0.08 & 33.49 $\pm$ 1.41 & 13.52 $\pm$ 1.02 \\
			\hline
		\end{tabular}
	\end{tabular*}
\end{table*}
It is reported in the literature that the temperature dependency behaviour of exchange bias and coercivity in FM/AFM and FM/SG systems is different~\cite{Wang-2018,Ali-2003}. Our motivation is to investigate the temperature dependency of EB in CoFeB/NiMn films and find out if it follows a trend similar to the FM/SG systems. The same experiments will also help us to find the blocking temperature of exchange bias. To perform temperature dependency of exchange bias, $\it{M-H}$ loops have been taken after FC from 400 K down to the desired temperatures (2, 10, 20, 30, 50, 100 and 300 K) in presence of 200 mT field. Figure 3 shows the plots of $H_{EB}$ and $H_{C}$ with temperature for all the samples and the experimental data were fitted using single exponential decay functions, given in eqs. (1) and (2), respectively, to find the presence of magnetic frustration in these systems ~\cite{Wang-2018,Ding-2013,Xie-2017}.
\begin{flalign}
H_{EB} (T) = H_{EB}^0 exp(-T/T_1)
\end{flalign}
\begin{flalign}
H_{C} (T) = H_{C}^0 exp(-T/T_2)
\end{flalign}
where $H_{EB}^0$ and $H_{C}^0$ are the exchange bias and coercive fields at 0 K, $T_{1}$ and $T_{2}$ are the constants. Table 2 shows the parameters obtained from the fit of the $H_{EB}$, $H_{C}$ vs $\it{T}$ experimental data using eqs. (1) and (2), respectively. 

It is reported in literature that the competition between Rudermann-Kittel-Kosuya-Yosida (RKKY) and direct exchange interactions gives the exponential decay of $H_{EB}$ and $H_{C}$ with temperature~\cite{Wang-2018}. RKKY interaction is the coupling of internal spins of SG and FM spins whereas direct exchange interaction is the coupling of surface spins of SG and FM spins~\cite{Wang-2018}. Thus, magnetic frustration gives exponential decay of $H_{EB}$ and $H_{C}$ with temperature. 

We found $T_{g}$ where the sudden rise of exchange bias occurs of $\sim$ 50 K for all the samples. The increase in number of frozen spins at low temperature region ($<$ 50 K) gives sudden rise in $H_{EB}$~\cite{Spizzo-2013,McCord-2013,Zhu-2015,Chandra-2015}. But, we also observed a sudden rise of $H_{C}$ at low temperature region ($<$ 50 K). Thus, at low temperature, the presence of large number of low anisotropy rotatable spins give a sudden rise in $H_{C}$~\cite{Spizzo-2013}. The spin glass like frustration might be a reason for the presence of both frozen and low anisotropy rotatable spins at temperature below 50 K~\cite{Spizzo-2013}. It is reported in literature that simultaneous decay of exchange bias field and coercive field w.r.t. temperature is not found in FM/AFM system~\cite{Ali-2003}. The temperature where exchange bias vanishes is known as blocking temperature $T_{B}$. We found that $T_{B}$ remains similar in all the sample. 
\subsection{Cooling field dependence of exchange bias}
\begin{figure*}
	\centering
	\includegraphics[scale=0.5]{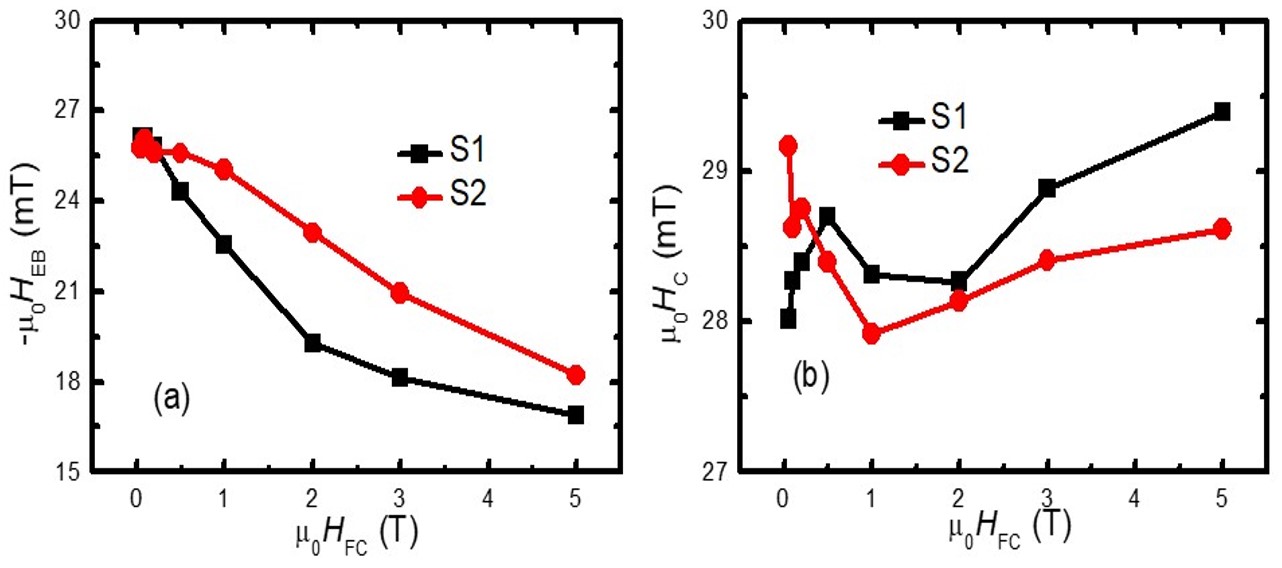}
	\caption{(a) $H_{EB}$ (b) $H_{C}$ vs $H_{FC}$ plots for the samples S1 and S2.}
	\label{fig:figure5}
\end{figure*}
In order to find additional evidence for the nature of interface, cooling field dependence of exchange bias can be performed. Figure 4 shows the trends of $H_{EB}$ and $H_{C}$ with $H_{FC}$ for the samples S1 and S2. To perform the cooling field dependency of EB, the samples were field cooled from 400 to 2 K in presence of various magnetic fields (0.05, 0.1, 0.2, 0.5, 1, 2, 3 and 5 T) and then $\it{M-H}$ loops were taken. In our study, we found the decrease of $H_{EB}$ with increase in cooling field whereas  $H_{C}$ remains almost constant. Similar trend of $H_{EB}$ and $H_{C}$ with cooling field has been reported in FM/SG system~\cite{Rui-2015}. However, in a FM/AFM system, $H_{EB}$ rises with increase in cooling field due to the enhancement in number of pinned moments along the cooling field direction~\cite{Bianco-2011}. We can interpret this behaviour as the presence of FM and AFM mix interactions in our systems. We found highest $H_{EB}$ in sample S2 at 10 K temperature. Thus, $H_{EB}$ is dependent on NiMn thickness indicating that not only the interface but also the $\lq${bulk}$\rq$  part of the NiMn contribute to exchange bias.
\subsection{Training effect}
\begin{figure*}
	\centering
	\includegraphics[scale=0.7]{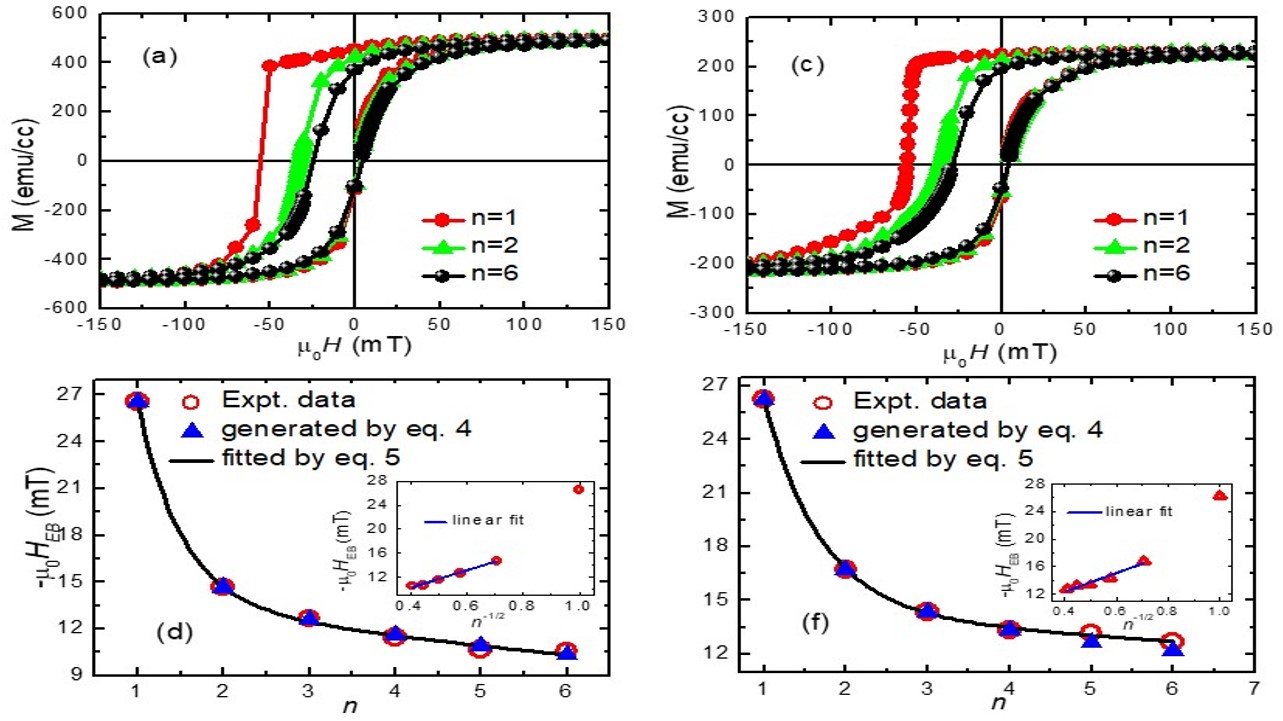}
	\caption{(a),(c) $1^{st}$, $2^{nd}$, and $6^{th}$ subsequent hysteresis loops of samples S1 and S2. (d),(f) solid circles are the experimental $H_{EB}$ vs $\it{n}$ data, blue triangular data points are generated from eq. (4) and solid line is the fitted data using eq. (5). The insets of the plots (d),(f) show the experimental $H_{EB}$ vs $n^{-1/2}$ data along with the fits using eq. (3) for the samples S1 and S2.}
	\label{fig:figure6}
\end{figure*}
One of the important properties  of the exchange bias systems is the training effect. Due to field cooling, the interfacial spins go to metastable states. However, consecutive cycling of the hysteresis loop without field cooling leads to relaxation of the metastable spins. The relaxation mechanism can be described by the training effect data analysis through various models.  We field cooled (FC) the samples down to 5 K from 400 K in the presence of 500 Oe field to record the first hysteresis loop and then the consecutive loops were taken for training effect measurements. Figure 5 (a),(c) show the  $1^{st}$, $2^{nd}$ and $6^{th}$ hysteresis loops and (d),(f) show the $H_{EB}$ vs $\it{n}$ data along with the fits using eqs. (4) and (5). The insets in figures 5 (d),(f) are the $H_{EB}$ vs $n^{-1/2}$ data with linear fits. We found large decrease in $H_{EB}$ in 2nd subsequent $\it{M-H}$ loop whereas gradual decrease is found after this (n $>$ 2). We could not fit the $H_{EB}$ vs $\it{n}$ data using thermal relaxation model as it excludes $\it{n}$ = 1~\cite{Shameem-2018}. However, to determine the value of $H_{EB\infty}$, we have fitted the $H_{EB}$ vs $n^{-1/2}$ data using the below equation~\cite{Paccard-1966};

\begin{flalign}
H_{EB} (n) = H_{EB\infty} + \frac{k}{n^{1/2}}
\end{flalign}
where $H_{EB}$(n) is the exchange bias field of $n^{th}$ loop, $H_{EB\infty}$ is the exchange bias field in the limit of infinite number of loops (n $\to$ $\infty$) and $\it{k}$ is the system dependent constant. 
The values of $H_{EB\infty}$ obtained from eq. (3) are given in table 3. We found that $H_{EB\infty}$ and $H_{EB}$ follow the similar trend. Power law decay of exchange bias has been observed in FM/AFM interfaces~\cite{Paccard-1966,Barman-2015,Shameem-2018}. But, in this study, spin glass like frustration is present.  Therefore, sudden decrease in $H_{EB}$ is found in $\it{n}$ = 2 loop due to less stability of interface spins under field reversal.

\begin{table*}
\small
	\caption{\label{tab:table1}The fitting parameters obtained using eqs. (3), (4) and (5).}
       	\begin{tabular*}{\textwidth}{@{\extracolsep{\fill}}ll}
		\begin{tabular}{cccccccc}
		 \hline 
			Sample name & $H_{EB\infty}$ (mT) & $\gamma_{H}$ ($10^{-3}$ $mT^{-2}$) & $A_{f}$ (mT) & $A_{i}$ (mT) & $P_{f}$ & $P_{i}$ & $P_{i}$/$P_{f}$ \\
			\hline
			S1 & 4.42 $\pm$ 0.49 & 1.88 & 93.48 $\pm$ 3.13 & 10.27 $\pm$ 0.48 & 0.50 $\pm$ 0.08 & 10.88 $\pm$ 1.36 & 21.76 \\
			S2 & 6.98 $\pm$ 0.83 & 2.55 & 51.61 $\pm$ 4.55 & 7.85 $\pm$ 0.68 & 0.67 $\pm$ 0.05 &  19.05 $\pm$ 6.19 & 28.43  \\	
    \hline 
    \end{tabular}
    \end{tabular*}
\end{table*}
\begin{figure*}[hbt!]
	\centering
	\includegraphics[scale=0.5]{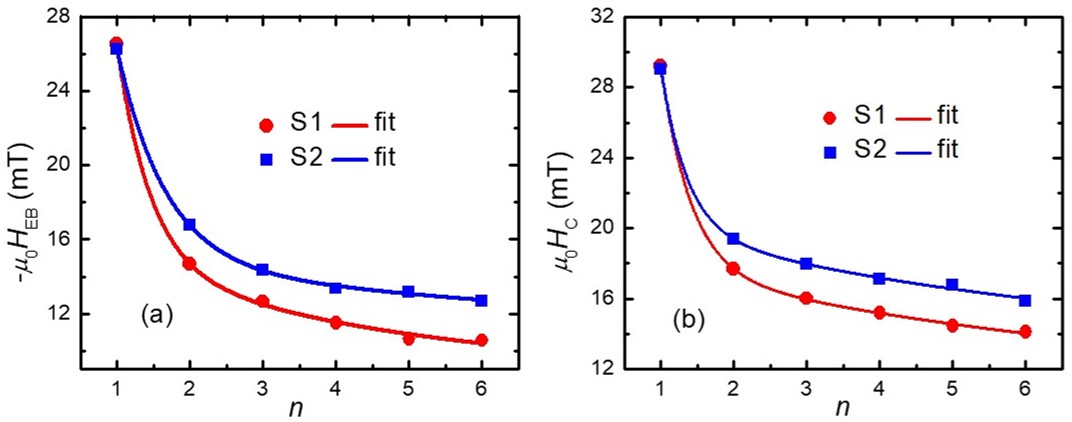}
	\caption{Solid symbols are the experimental data of (a) $H_{EB}$ and (b) $H_{C}$ vs $\it{n}$ for all the samples with the solid lines are the fits using eq. (6) for the samples S1 and S2.}
	\label{fig:figure7}
\end{figure*}
\begin{figure*}[hbt!]
	\centering
	\includegraphics[scale=0.6]{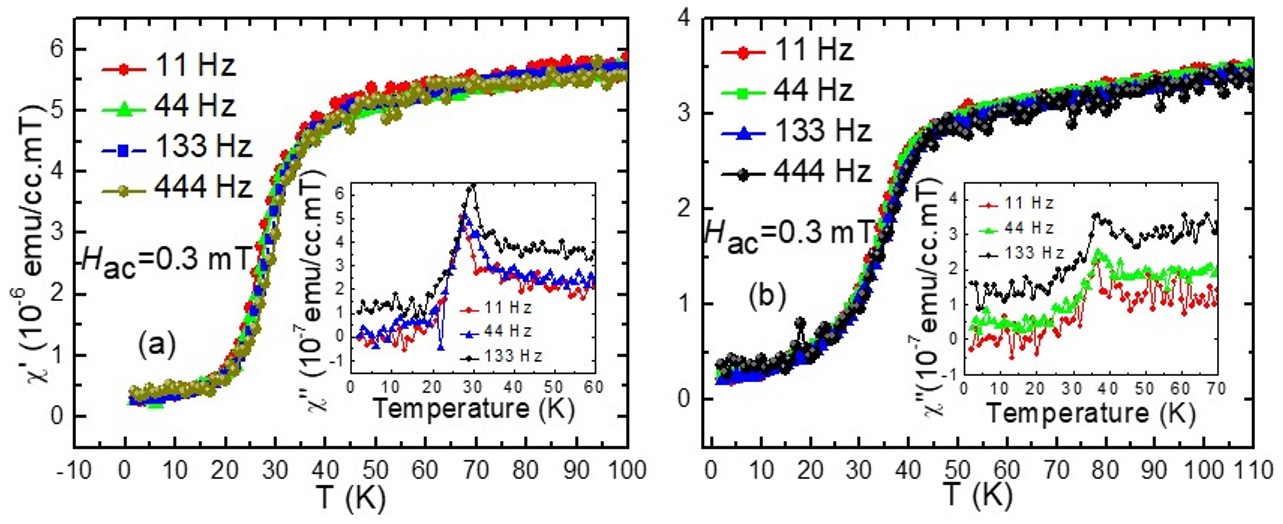}
	\caption{(a),(b) Plots of real part of ac susceptibility ($\chi{^\prime}$) vs temperature ($\it{T}$) and insets show the imaginary part of ac susceptibility ($\chi^{\prime \prime}$) vs temperature ($\it{T}$) plot at different frequencies for the samples S1 and S2. }
	\label{fig:figure8}
\end{figure*}
As eq. (3) failed to explain the training effect data, We considered the model given by Binek which is known as spin configurational relaxation model and is given below~\cite{Binek-2004}; 

\begin{flalign}
H_{EB} (n+1)-H_{EB} (n) = -\gamma_{H} (H_{EB} (n)-H_{EB\infty})^{3}
\end{flalign}

where $H_{EB}$(n), $H_{EB}$(n+1) and $H_{EB\infty}$ are the exchange bias fields of the $n^{th}$, $(n+1)^{th}$ and in the limit of infinite number of loops, respectively. $\gamma_{H}$ is the characteristic decay rate of the training effect and defined as $\gamma_{H}$=b/($K^{2}$$\zeta$) where $\it{K}$ is a constant proportional to the exchange coupling strength in FM/AFM system, $\zeta$ is the inverse of relaxation time and $\it{b}$ is another constant~\cite{Shameem-2018}. The small value of $\gamma_{H}$ indicates large deviation from equilibrium steady state and hence large training effect which is according to relaxation theory~\cite{Wang-1966}. Such spin configurational relaxation in training effect data is also reported for a FM-SG interface. 

\begin{table*}[hbt!]
	\caption {The fitting parameters obtained utilizing eq. (6).}
	\footnotesize
	\begin{tabular}{p{5em}p{4em}p{4em}p{4em}p{4em}p{4em}p{5em}p{5em}p{3em}p{3em}p{3em}p{3em}p{3em}}
		Sample name & $H_{E}$($\infty$) (mT) & $A_{S}^{EB}$ (mT) & $A_{i}^{EB}$ (mT) & $n_{0}^{EB}$ & $\tau_{S}^{EB}$ & $\tau_{i}^{EB}$ & $H_{C}$($\infty$) (mT) & $A_{S}^{CO}$ (mT) & $A_{i}^{CO}$ (mT) & $n_{0}^{CO}$ & $\tau_{S}^{CO}$ & $\tau_{i}^{CO}$ \\ \hline
	    	S1 & 8.52  $\pm$0.21 & 12.31  $\pm$1.05 & 6.13  $\pm$0.21 &  0.98  $\pm$0.20 & 0.45  $\pm$0.09 & 4.57$\pm$ 0.98 & 10.21$\pm$0.09 & 12.11  $\pm$ 0.09 & 7.35   $\pm$ 0.35 & 0.98  $\pm$0.06 & 0.41  $\pm$0.03 & 7.65  $\pm$0.75 \\
			S2 & 8.71  $\pm$0.05 & 11.76  $\pm$0.15 & 5.78  $\pm$0.09 & 1.00  $\pm$0.03 & 0.67  $\pm$0.06 &  13.57  $\pm$1.12 & 12.86  $\pm$0.31 & 9.06  $\pm$0.15 & 6.89  $\pm$0.09 & 1.00  $\pm$0.08 & 0.36  $\pm$0.03 & 6.32  $\pm$0.22  \\	
      	\end{tabular}\\
\end{table*}
Another approach to explain training effect data was given by Mishra et al. which considers relaxation rate of both the rotatable and frozen spins at the interface and is described below~\cite{Mishra-2009};

\begin{flalign}
H_{EB} (n)=H_{EB\infty}+A_{f}exp(-n/P_{f})+A_{i}exp(-n/P_{i})
\end{flalign}

where $A_{f}$, $P_{f}$ are the interfacial frozen spin parameters and $A_{i}$, $P_{i}$ are the interfacial rotatable spin parameters of the NiMn/CoFeB system. $P_{f}$ and $P_{i}$ are the relaxation rates of interfacial frozen and rotatable spins, respectively. $P_{f}$ and $P_{i}$ are dimensionless. $A_{f}$ and $A_{i}$ have the dimension of mT. We observed that $A_{f}$ and $A_{i}$ are decreasing with the thickness of NiMn. Thus, the frozen and rotatable interfacial spin components become lesser with increase in thickness of NiMn. Also, $A_{f}$ is higher than $A_{i}$ indicating that the frozen spin components have major contribution to the training effect. $P_{f}$ is almost constant for all the samples. However, $P_{i}$ increases as NiMn becomes thicker leading to an increment in $P_{i}$/$P_{f}$ indicating that not only the interface but also $\lq${bulk}$\rq$ of NiMn contributes to the relaxation. 

The model given by Binek considers only instability of interface AFM magnetization whereas Mishra et al. introduced the relaxation of frozen and rotatable interface spins. 
Parameters obtained by the fits using eqs. (3), (4) and (5) are given in table 3.
In the training effect $\it{M-H}$ loops (figure 5), the magnitude of $H_{EB}$ reduces in the descending part of the loop whereas the magnitude remains constant in the ascending part of the loop similar to FM/AFM systems~\cite{Yuan-2011}. 

Above models describe the training induced relaxation of interface magnetization. However, we want to confirm that not only FM/SG interface but also $\lq${bulk}$\rq$ spins of SG contribute to relaxation in training effect. In order to separate the contribution of the $\lq${bulk}$\rq$ NiMn and interface spins of NiMn/CoFeB system towards training induced relaxation, we fitted the training effect data using the following eq.~\cite{Chi-2016};
\begin{flalign}
\pm H_{E,C}(n) & = \pm H_{E,C}(\infty) \nonumber\\
& + A_{S}^{EB,CO} exp[- (n-n_{0}^{EB,CO})/\tau_{S}^{EB,CO}] \nonumber\\
& + A_{i}^{EB,CO} exp[- (n-n_{0}^{EB,CO})/\tau_{i}^{EB,CO}] \nonumber\\
\end{flalign}

where  $H_{E}$($\infty$) and $H_{C}$($\infty$) are the limiting values of exchange bias and coercive fields after infinite number of loop run (n $\to$ $\infty$). $A_{s}^{EB,CO}$ and $A_{i}^{EB,CO}$ have the dimension of magnetic field in which the superscripts EB, CO correspond to the exchange bias and coercivity whereas the subscripts $\it{s}$, $\it{i}$ indicate the weights of spin glass $\lq${bulk}$\rq$ and spin-glass-like interface, respectively. $\it{n}$, $n_{0}$ and $\tau$ have the dimensions of time. Relaxation velocity is determined by $\tau$ and the relaxation will be faster for smaller $\tau$. $n_{0}$ is the shifting coefficient. Figure 6 shows the $H_{EB}$, $H_{C}$ vs $\it{n}$ data for all the samples along with the fits using eq. (6). Table 4 shows the parameters obtained by the fits using eq. (6). 
The amplitude of decay $A_{i}^{EB}$ for the spin-glass-like interface is found to be dependent on NiMn thickness whereas $A_{s}^{EB}$ which is the amplitude of decay of $\lq${bulk}$\rq$ NiMn spins is independent on the NiMn thickness. It is found that the magnitude of $A_{s}^{EB}$ is higher than $A_{i}^{EB}$ in all the samples. Similarly, the magnitude of  $A_{s}^{CO}$ is higher than $A_{i}^{CO}$. In all samples, $\tau_{i}^{EB}$ has a greater magnitude than $\tau_{s}^{EB}$ indicating that the $\lq${bulk}$\rq$ spins of NiMn spin glass relax faster than spin-glass-like interface spins. Thus, for smaller $\it{n}$, $\lq${bulk}$\rq$ NiMn spins play dominant role whereas the spin-glass-like interface spins play role for higher $\it{n}$ in training induced relaxation. The relaxation rate of spin-glass-like interface spins $\tau_{i}^{EB}$ is dependent on the NiMn thickness. It is also reported that the value of shifting coefficient $n_{0}$ is approx. 1 in FM/SG system. We also found the value of $n_{0}$ $\sim$ 1 in this study. 

\begin{table*}[hbt!]
	\caption{\label{tab:table 2}Fitting parameters obtained using eqs. (7) and (8) of all the samples.}
		\begin{tabular}{cccccc}
			&\multicolumn{2}{c}{N-A Model}&\multicolumn{3}{c}{V-F law}\\
			Sample name&$f_{0}$ (Hz)&$E_{a}$/$K_{B}$ (K)&$\tau_{0}$ (s)&  $E_{VF}$/$k_{B}$ (K) & $T_{VF}$ (K) \\ \hline
			S1&2.44$\times$$10^{16}$$\pm$7.22$\times$$10^{14}$&951.88$\pm$111.90&4.66$\times$$10^{-12}$$\pm$6.91$\times$$10^{-13}$&657.73$\pm$53.12&0.20$\pm$0.10 \\
			S2&5.88$\times$$10^{19}$$\pm$4.22$\times$$10^{18}$&1461.04$\pm$2.92&3.19$\times$$10^{-12}$$\pm$3.26$\times$$10^{-13}$&438.74$\pm$83.32&17.29
			$\pm$1.71 \\
		\end{tabular}
\end{table*}
\subsection{AC susceptibility}

\begin{figure*}[hbt!]
	\centering
	\includegraphics[scale=0.6]{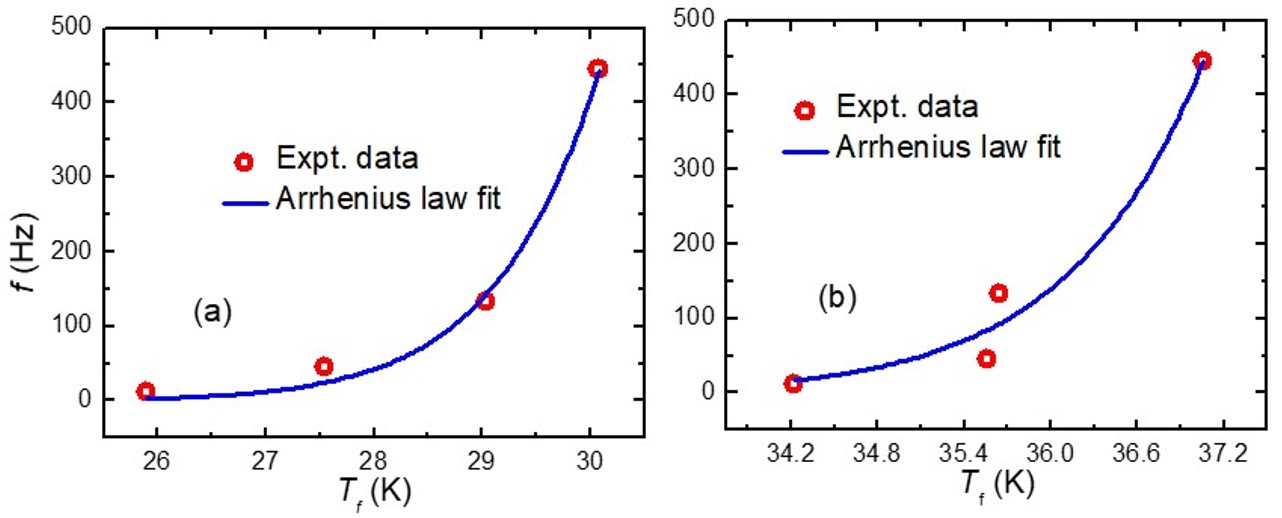}
	\caption{(a),(b) Plots of frequency ($f$) vs $T_{f}$ experimental data with fitted data using N-A model for the samples S1 and S2.}
	\label{fig:figure9}
\end{figure*}
\begin{figure*}[hbt!]
	\centering
	\includegraphics[scale=0.6]{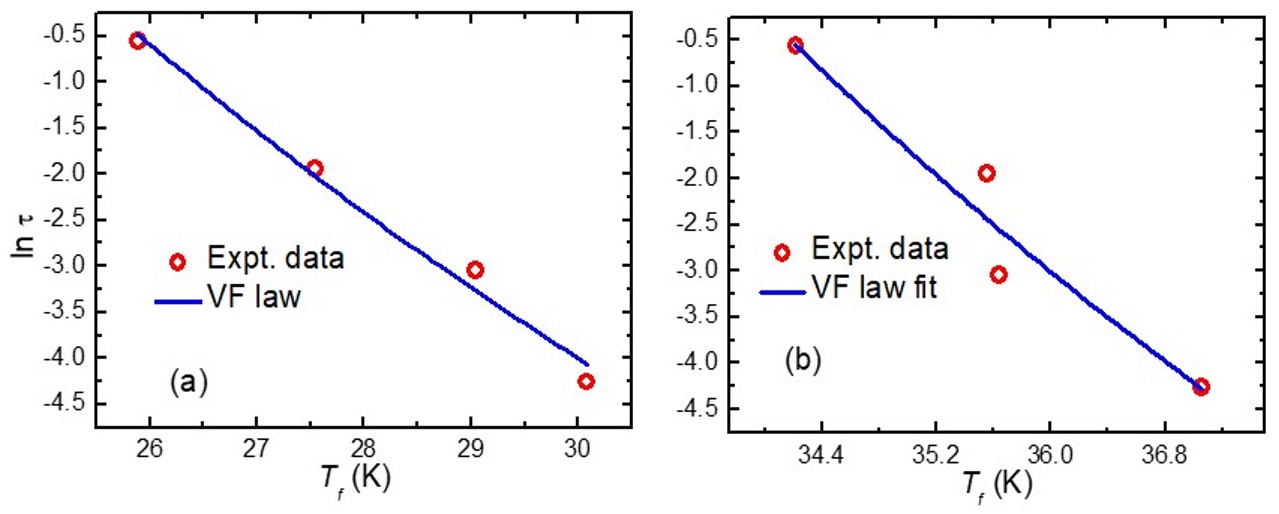}
	\caption{(a),(b) Plots of ln($\tau$) vs $T_{f}$ experimental data with fitted data using V-F law for the samples S1 and S2.}
	\label{fig:figure10}
\end{figure*}
To provide additional evidence about the magnetic nature of NiMn in the NiMn/CoFeB bilayer, we performed ac susceptibility measurements w.r.t. temperature at different frequencies in presence of an ac field of 0.3 mT. Figure 7 shows the plots of real part of ac susceptibility ($\chi^\prime$) vs temperature and the insets show the plots of imaginary part of ac susceptibility ($\chi^{\prime \prime}$) vs temperature at different frequencies for all the samples. We found that the peak temperature $T_{f}$, obtained from the real part of the ac susceptibility ($\chi^\prime$) vs temperature ($\it{T}$) plot, shifts towards higher temperature confirming the SG nature of NiMn. 
$T_{f}$ has shifted from $\sim$ 26 K at 11 Hz to $\sim$ 30 K at 444 Hz in sample S1. In sample S2, $T_{f}$ gets shifted from $\sim$ 34 K at 11 Hz frequency to $\sim$ 37 K at 444 Hz frequency. 
The dynamic behaviour of magnetic system is governed by temperature also~\cite{Slimani-2018}. 
Neel-Arrhenius (N-A) proposed a model which considers both the anisotropy energy $E_{a}$=$K_{eff}$V and thermal energy $k_{B}$T. $K_{eff}$ is the effective anisotropy constant which takes into account surface, interface anisotropies etc. and V is the volume of particles. N-A model can be expressed as follows~\cite{Slimani-2018}; 
\begin{flalign}
f = f_{0}  exp (-E_{a}/K_{B}T_{f})
\end{flalign}
where, $f$ is the rate of flipping of magnetization between the two lowest energy states, an attempt frequency is defined as $f_{0}$ whose value for superparamagnets lies between $10^{8}$ to $10^{12}$ Hz~\cite{Tiwari-2005,Labarta-1993}. $k_{B}$ is the Boltzmann constant. 
The values of $f_{0}$ are found to be larger than the usual values of the superparamagnets. We obtained unphysical large $E_{a}$/$k_{B}$ values from the fit using  N-A model~\cite{Tiwari-2005}. The plots of frequency ($\it{f}$) vs $T_{f}$ are shown in figure 8. It is assumed that the presence of interactions tune $T_{f}$ through the modification of energy barrier. Thus, the anisotropy energy $E_{a}$ and relaxation time are tuned not only by thermal energy but also by the interaction present in the system~\cite{Slimani-2018}. To explain magnetically interacting system, Vogel-Fulcher (V-F) described a theory~\cite{Bedanta-2009,Chandrasekhar-2012,Slimani-2018}; 
\begin{flalign}
\tau = \tau_{0} exp (E_{VF}/K_{B}(T_{f}-T_{VF}))
\end{flalign}
Where, $E_{VF}$ is the activation energy and Vogel-Fulcher temperature, $T_{VF}$, is a measure of the interaction strength. 
The value of $\tau_{0}$ for a spin glass or cluster spin glass system lies in-between $10^{-12}$ to $10^{-14}$ s~\cite{Chandrasekhar-2012}. Figure 8 shows the fit of $T_{f}$ vs ln$\tau$.
We found the values of $\tau_{0}$ similar to spin glass systems. The fitting parameters obtained using eqs. (7) and (8) are given in table 5.

In summary, presence of magnetic frustration can be concluded from the exponential decay of both $H_{EB}$ and $H_{C}$ with temperature, the cooling field dependence of exchange bias, ac susceptibility measurements etc. We found the blocking temperature $T_{b}$, where maximum number of particles are unblocked, from the $\it{M-T}$ measurements. Sudden rise in $H_{EB}$ is found below $\sim$ 50 K from temperature dependence of exchange bias due to the role of frozen spins of SG. Again, the decrease in exchange bias field $H_{EB}$ is observed with the increase in cooling field $H_{FC}$. We fitted the training effect data using various models. Among them, thermal relaxation model fails to explain training effect. We have investigated from the training effect fitting that not only the interface but also $\lq${bulk}$\rq$ NiMn spins contribute for the relaxation. The relaxation time $\tau_{0}$ obtained from V-F law fitting indicates that the system has non-negligible interaction like spin glass.
\section*{Conflicts of interest}
There are no conflicts to declare.
\section*{Acknowledgments:}
The authors thank Department of Atomic Energy (DAE) for providing the financial support. BBS acknowledges DST for INSPIRE faculty fellowship. We  acknowledge Pushpendra Gupta for his help in SQUID measurements.

\end{document}